\documentclass[preprint, 12pt]{elsarticle} 

\usepackage{graphicx,amssymb}
\usepackage[usenames,dvipsnames]{color}
\usepackage{epsfig}
\usepackage{overpic}

\newcommand{\eqref}[1]{(\ref{#1})}

\begin{document}
 
\begin{frontmatter}
\title{
Dynamical response of ultracold interacting fermion-boson mixtures
}
\author[]{Kai Ji$^{a,b}$}
\author[]{Stefan Maier$^{a}$}
\author[]{Andreas Komnik$^{a,c,}$\footnote{Corresponding author, Institut f\"ur Theoretische Physik, Universit\"at Heidelberg, Philosophenweg 12, 69120 Heidelberg, Germany; Email: komnik@uni-heidelberg.de, Phone +49 (0)6221 54  5049, Fax  +49  (0)6221 54 9331 }}
\address[]{Institut f\"ur Theoretische Physik, Universit\"at Heidelberg, Philosophenweg 12, D-69120 Heidelberg, Germany}
\address[]{{Theory of Quantum and Complex Systems (TQC), Universiteit Antwerpen, Universiteitsplein 1, B-2610 Antwerpen, Belgium}}
\address[]{Freiburg Institute of Advanced Studies (FRIAS), Universit\"at Freiburg, Albertstr. 19, D-79104 Freiburg i. Br., Germany}

\date{\today}

\begin{abstract}
 We  analyze the dynamical response of a ultracold binary gas mixture in presence of strong boson-fermion couplings. Mapping the problem onto that of the optical response of a metal/semiconductor electronic degrees of freedom to electromagnetic perturbation we calculate the corresponding dynamic linear response susceptibility in the non-perturbative regimes of strong boson-fermion coupling using diagrammatic resummation technique as well as quantum Monte Carlo simulations. We evaluate the Bragg spectral function as well as the optical conductivity and find a pseudogap, which forms in certain parameter regimes. 
\end{abstract}
\end{frontmatter}

\section{Introduction}
\label{sec:intro}

Ultracold gas mixtures of bosons and fermions, which arise very naturally in the sympathetic cooling processes, are very interesting quantum systems with physical properties very different from conventional quantum gases \cite{Truscott30032001,Modugno27092002}. 
In view of recent advances in the field of correlated ultracold gases it is very important to understand their dynamical properties, e.~g. their response to an external dynamical scattering potential brought about by a change of trapping potential. If the bosonic subsystem is dominantly in the BEC phase the effective low-energy interaction with the fermionic subsystem is the coupling between the latter and the phonons (sound waves) of the former. From the mathematical point of view such a system is nothing but an electron-phonon coupled system best described by the Fr\"ohlich Hamiltonian{\color{black}, see Eq.~\eqref{NRone} in Section \ref{sec:model}, the interaction term of which describes precisely the scattering of fermions on the bosonic degrees of freedom mentioned above} \cite{Froehlich}. In {\color{black} the} case of low fermion concentrations it describes individual impurities imbedded into a continuum of massless bosonic modes. Under such conditions the physics of the system is supposed to be very close to that of the classical polaron, taking place in semiconductors with strong electron-phonon interaction \cite{LandauPekar,PhysRev.97.660,Tempere_0}.

There are, however, fundamental differences between the conventional (solid state) polarons and their BEC counterparts. The most obvious one is the different phonon spectrum of the bosonic subsystem as well as an explicit momentum dependence of the electron-phonon coupling \cite{Tempere_0}. While these details do not alter the general picture of polaron static properties (there is still an effective mass generation and self-trapping), they could possibly alter the dynamical response, which reveals such important information as how the impurities interact with their surroundings \cite{Goodvin2011,Hohenadler2005a}. In this paper we would like to consider them in full detail and in different geometries with the special emphasis on strong coupling results thereby closing the gaps in the existing literature.  

The paper is organized as follows. In the next section we formulate the problem and introduce all relevant quantities. Section \ref{sec:RPA} is devoted to the non-perturbative approach inspired by the classical random phase approximation (RPA). We explain the details of the implementation and discuss the special features pertinent to ultracold gas realizations. In Section \ref{sec:MC} the calculation of the dynamical response function is accomplished using numerically exact quantum Monte Carlo (QMC) simulation technique. Section \ref{sec:discussion} contains a discussion of results and offers several avenues of further progress.

\section{The model and observables}
\label{sec:model}

An effective low-energy Hamiltonian for a BEC-fermion mixture has the canonical Fr\"ohlich form \cite{Froehlich,Tempere_0}, which is written in terms of boson (described by the field operators $b_{\bf k}$) and fermion (denoted by $a_{\bf q}$) degrees of freedom,
\begin{eqnarray}                                    \label{NRone}
  H &=& \sum_q (E_q -\mu) \, a_{\bf q}^\dag a_{\bf q} + \sum_k \omega_k b_{\bf k}^\dag b_{\bf k} 
  \nonumber \\
  &+&
  \sum_q \sum_{k\neq 0} V_{\bf k} \, a^\dag_{\bf q + k} a_{\bf q} \, (b_{\bf k} + b^\dag_{- \bf k}) \, ,
\end{eqnarray}
where the dispersion of fermions is $E_q = q^2/2m$, $\mu$ is their chemical potential, 
\begin{eqnarray}       \label{phonondisp}
\omega_k = c k [1 + (\xi k)^2/2]^{1/2}
\end{eqnarray}
is the dispersion of the phonon mode
{with an effective mass $m_p$, $c = (\sqrt{2} m_p \xi)^{-1}$ is the speed of sound in the condensate,}
and the coupling is given by $V_{\bf k} = \lambda [(\xi k)^2/ ((\xi k)^2 + 2)]^{1/4}$ with $\lambda = g_{\rm IB} \, \sqrt{N_0}$. $g_{\rm IB}$ is the effective interaction strength between {\color{black} the} impurities and Bogoliubov excitations and can be adjusted by changing the particle density and/or the $s$-wave scattering length of collision processes of the impurity with the bosonic medium. $\xi$ denotes the healing length of the condensate and is given by $\xi=1/\sqrt{8\pi a_{BB}
 n_0}$ where $a_{BB}$ is the boson-boson $s$-wave scattering length and $n_0$ is the condensate density. 
We would like to point out that the model  \eqref{NRone} is valid for not too strong boson-fermion scattering. As soon as the Bogolyubov approximation breaks down one has to work with the full interacting Hamiltonian \cite{Tempere_0}. 
Nonetheless, it was demonstrated previously, that realistic boson-fermion mixtures turn out to show many details, which are adequately described by the strong coupling limit of the much simpler Fr\"ohlich Hamiltonian \cite{Cucchietti_Timmermans}. That is why we concentrate on \eqref{NRone} throughout the paper.
Yet another issue is that strong interactions might change the condensate fraction and thus 
influence the system parameters. As we focus on not too strong interactions we would like to neglect these effects.

One fundamental difference between the `classical' semiconductor based electron-phonon coupled system and the one realized in ultracold mixtures is that the quantum gas system can be prepared in trapping potentials for fermions and bosons which might be of different shape and dimensionality. For that reason we shall later consider systems with different $E_{\bf q}$ and $\omega_{\bf k}$. Changing the shape of the trapping potential for the impurity in space and time, for instance by acceleration with respect to the BEC, which rests in the laboratory reference frame one induces the rearrangement of particles. In this way one can access the mobility of the impurity, which, like in the case of a Brownian motion, is the ultimate dynamical quantity of the particle \cite{Kubo}.
In detail, the mobility is found from the velocity autocorrelation function, which translates into the current-current correlation function 
$\Pi\left({\bf q},\omega\right)$ in the Matsubara representation,
\begin{eqnarray}   \label{momdepcond}
\Pi_{\mu\nu}\left({\bf q},\tau\right) = -\frac{1}{V}\left\langle T_\tau j_\mu^\dag\left({\bf q},\tau\right) j_\nu\left({\bf q},0\right)\right\rangle \, ,
\end{eqnarray}
with the current densities defined by
\begin{eqnarray}
{ {\bf j}} \left({\bf q}\right) = - \frac{1}{m \beta V} \sum_{\bf k}\left({\bf k} + \frac{\bf q}{2}\right)a^\dag_{{\bf k} + {\bf q}} a_{\bf k}\, .
\end{eqnarray}
This picture is very similar to the conventional polaron problem in semiconductors, where the principal quantity is the \emph{momentum-dependent optical conductivity} \cite{bruus2004book},
\begin{eqnarray}
\mbox{Re}[\sigma_{\mu\nu}\left({\bf q},\omega\right)] = -\frac{e^2}{\omega}\mbox{Im}[\Pi_{\mu\nu}^{R}\left({\bf q},\omega\right)] \, .
\end{eqnarray}
Here {by abuse of notation} the subscript $\mu$ in the double sum indicates the spatial direction with respect to which the conductance is probed, that is $\mu\in\left\{x,y,z\right\}$.
The superscript $R$ denotes the retarded correlation function, which is obtained from the one in the Matsubara representation by the usual analytic continuation.
From the perspective of a solid-state physicist, the optical conductivity computed at ${\bf q}=0$ describes the experimental conditions quite well, that is probing a sample with optical or X-ray photons does not lead to a substantial momentum transfer ($\Delta {\bf p} \approx 0$). In case of the RF-spectroscopy or the aforementioned experimental procedures in ultracold quantum gases, this is not necessarily the case. That is why throughout the paper we shall consider both ${\bf q}=0$ and finite ${\bf q}$ situations whenever possible.

Another experimental technique to access the impurity dynamics is the Bragg spectroscopy
\cite{RevModPhys.80.885,PhysRevLett.93.080401,PhysRevLett.82.4569,Bragg1,Bragg2,Casteels2011a}. {\color{black} In a typical measurement the BEC is subject to two noncollinear laser beams with photons with wave vectors ${\bf k}_{1,2}$ and energies $\omega_{1,2}$. The fermionic atoms can then undergo a stimulated scattering absorbing the light from the beam 1 and emitting it into the laser field 2, thereby acquiring momentum and energy given by the differences of ${\bf k}_{1,2}$ and $\omega_{1,2}$. How much of the momentum and energy is absorbed by the BEC can then be mapped out by time-of-flight measurements after the trap release \cite{Casteels2011a}. The absorption spectra are then directly related to the autocorrelation of the particle density (here ${\bf q}={\bf k}_{1}-{\bf k}_{2}$): }
\begin{eqnarray}                          \label{autodensity}
\chi ({\bf q},\tau) = - {1 \over V} \langle T_{\tau} \rho^{\dag} ({\bf q},\tau) \rho ({\bf q},0) \rangle \, ,
\end{eqnarray}
where 
\begin{eqnarray}
\rho({\bf q})= \sum_{\bf k} a_{{\bf k} + {\bf q}}^{\dag} a_{\bf k}\, , 
\end{eqnarray}
is the particle density operator. Very conveniently the \emph{optical absorption spectrum}
\begin{eqnarray}
R_{\Pi}^{\mu \nu} ({\bf q}, \omega) = -{1 \over \pi} \mbox{Im} \, \Pi_{\mu \nu}^R ({\bf q}, \omega) \, , 
\end{eqnarray}
and \emph{Bragg spectrum} (or Bragg spectral function), which we define as
\begin{eqnarray}
R_{\chi} ({\bf q}, \omega) = -{1 \over \pi} \mbox{Im} \, \chi^R ({\bf q}, \omega) \, , 
\end{eqnarray}
are related to each other in the following way (see \ref{sec:app_spec}):
\begin{eqnarray}                \label{fund_relation}
R_{\chi} ({\bf q}, \omega) &=& \left( {q \over \omega e} \right)^2 R_{\Pi}^{\parallel} ({\bf q}, \omega) \, ,
\end{eqnarray}
where `$\parallel$' specifies the component in the direction of $\bf q$.
Thus, once the current autocorrelation function is computed we have access to measurable quantities for both experimental schemes. 

Now we would like to translate the autocorrelation function \eqref{momdepcond} into the operator language of the original Hamltonian. From now on we skip the vector notation since we would like to restrict ourselves to 1D only. Experimentally, this can be motivated by the use of quantum gases in reduced dimensions. {Although in translationally invariant 1D systems no BE condensation is possible, in a realistic experimental situation there is always a confinement potential which facilitates a condensation. That is why it is legitimate to work in that picture.}
In energy-momentum representation
we {then} obtain
\begin{eqnarray}
\Pi (q, i \omega_n) = -{1 \over V} \int_0^{\beta} d \tau e^{i \omega_n \tau}
\langle T_\tau j^{\dag} (q, \tau) j (q,0) \rangle
\end{eqnarray}
\begin{eqnarray}                               \label{longformula_1}
& = & \frac{e^2 q^2}{(i \omega_n)^2 m^3 V} \sum_k \left(3k^2 + {q^2 \over 4} \right) \langle a_k^{\dag} a_k \rangle
- \frac{e^2}{(i \omega_n)^2 m^2 V} \sum_{q'} q' (q + q') V_{q'}^* \langle B_{q'} \rho^{\dag} (q') \rangle
\nonumber \\
&-& \frac{e^2}{(i \omega_n)^2 m^2 V} \int_0^{\beta} d\tau e^{i \omega_n \tau}
\left[ \frac{q^2}{m^2} \sum_{kk'} \left( k + {q \over 2} \right)^2 \left( k' + {q \over 2} \right)^2
\langle T_\tau a_k^{\dag} (\tau) a_{k+q} (\tau) a_{k'+q}^{\dag} a_{k'} \rangle
\right. \nonumber \\
&-& \left. \frac{q}{m} \sum_{kq'} V_{q'} q' \left( k + {q \over 2} \right)^2
\langle T_\tau B_{q'}^{\dag} a_k^{\dag} (\tau) a_{k+q} (\tau) \rho (q+q') \rangle
\right.
\nonumber \\
&-& \left. \frac{q}{m} \sum_{kq'} V_{q'}^* q' \left( k + {q \over 2} \right)^2
\langle T_\tau B_{q'} \rho^{\dag} (q+q', \tau) a_{k+q}^{\dag} a_k \rangle
\right.
\nonumber \\
&+& \left. \sum_{q'q''} V_{q'}^* V_{q''} q' q''
\langle T_\tau B_{q'} (\tau) B_{q''}^{\dag} \rho^{\dag} (q+q', \tau) \rho (q+q'') \rangle \ \right] \, , 
\end{eqnarray}
where $B_q = b_q + b_{-q}^\dag$.  This formula will be evaluated in the analytical computations of the following section, while for the QMC simulation of Section \ref{sec:MC} the current and density correlations would be computed in a slightly different way.

If not explicitly stated otherwise we use the polaronic units. Distances are measured in {units of} $\xi$, time in units of $m \xi^2/\hbar$ and energies in units of $\hbar^2/(m \xi^2)$. 
{In numerical calculations in Section \ref{sec:MC}, we introduce the specific system of $^6$Li impurities in a BEC of $^{23}$Na which renders $m_p/m=3.8$ \cite{Tempere_0}.}

\section{RPA results for the response function}
\label{sec:RPA}

Analytical approaches have been widely used in the investigation of the effects of electron-phonon interactions in metals and semiconductors (for the latest review, consult e.~g. Ref.~\cite{Devreese2009rev}). Very recently, due to formal similarities much effort has been made to adapt these methods to Bose-Fermi mixtures \cite{Bruderer2007a, novikov2010a, Tempere_0, Casteels2011a, Casteels2011b, Casteels2011c, Casteels2011d, Mathy:2012fk}. Although the applicability and accuracy of these approximation schemes have in general to be questioned \cite{Mishchenko2007collect}, their simplicity allows for an easy way to get first insights into the complex nature of electron-phonon interaction. Especially when it comes to the interpretation of numerical data, such analytical models proved to be of much use and importance.

We begin our calculations with a simple perturbative treatment of the 
optical absorption spectra and Bragg spectral function. We start with the current autocorrelation function $\Pi(q,\omega)$. It turns out that in the lowest order of boson-fermion coupling strength only the third and the sixth terms of \eqref{longformula_1} contribute. After the analytic continuation procedure we obtain the following result:
\begin{eqnarray}                               \label{longformula_2}
\Pi (q, \omega) = & & \frac{e^2 q^2}{\omega^2 m^4} \int {dk \over 2 \pi} \left( k + {q \over 2} \right)^2 n_F (\epsilon_k) \left[1 - n_F (\epsilon_{k+q}) \right]
\nonumber \\
& & \times \left[ P \left( \frac{1}{\omega + \epsilon_k - \epsilon_{k+q}} \right) - i \pi \delta (\omega + \epsilon_k - \epsilon_{k+q}) \right]
\nonumber \\
& & \times \left[ 1 - e^{\beta (\epsilon_k - \epsilon_{k+q})} \right] - \frac{e^2}{\omega^2 m^2} \int {dq' \over 2 \pi} |V_{q'}|^2 q'^2
\nonumber \\
& & \times \int {d \varepsilon \over \pi} \{ n_B (\varepsilon) [ \mbox{Im} \chi_0^R (q+q',\varepsilon)  \mbox{Re} D_0^R (q',\varepsilon-\omega)
\nonumber \\
& & + \mbox{Im} D_0^R (q',\varepsilon) \mbox{Re} \chi_0^R (q+q',\varepsilon+\omega) - i [ n_B(\varepsilon+\omega)
\nonumber \\
& & - n_B (\varepsilon) ] \mbox{Im} D_0^R (q',\varepsilon) \mbox{Im} \chi_0^R (q+q',\varepsilon+\omega) \} \, ,
\end{eqnarray}
where $n_B$ denotes the Bose distribution function and $\chi_0^R (q,\omega)$ is the retarded density correlation function \eqref{autodensity} of the non-interacting boson-fermion mixture, given by
\begin{eqnarray}
\chi_0 (q, i \omega_n) = \frac{1}{V}\sum_k \frac{n_\mathrm{F}\left(\epsilon_k\right)-n_\mathrm{F}\left(\epsilon_{k+q}\right)}{i \omega_n + \epsilon_k - \epsilon_{k+q}} \, .
\end{eqnarray}
$n_\mathrm{F}$ denotes the Fermi-Dirac distribution function and $\epsilon_k = E_k - \mu$ with the chemical potential $\mu$. 
$D^R (q,\nu)$ is the retarded Green's function of the {phonon}, which is obtained from the conventional Matsubara definition:
\begin{eqnarray}
 \mathcal{D}(q,\tau) = \langle T_\tau B_q(\tau) B_q^\dag \rangle \, 
\end{eqnarray}
via standard procedure. For the free case it is obviously
\begin{eqnarray}
{\mathcal D}^{\left(0\right)}\left(q,i\nu_n\right) = \frac{1}{i\nu_n - \omega_q} -\frac{1}{i\nu_n + \omega_q}
\, .
\end{eqnarray}
Putting everything together into \eqref{longformula_2} we obtain the following result for the Bragg spectral function:
\begin{eqnarray}                               \label{Bragg_pert} 
&& R_{\chi} (q,\omega) = {m \over 2 \pi |q|} n_F (\epsilon_p)|_{p={m \omega \over q} - {q \over 2}} \left[ 1 - n_F (\epsilon_p)|_{p={m \omega \over q} + {q \over 2}} \right]
\nonumber \\
&\times& \left( 1 - e^{-2 \beta m \omega} \right)
    + \frac{q^2}{2 \pi m \omega^4} \int {dq' \over 2 \pi} |V_{q'}|^2
 \\ \nonumber
&\times& {q'^2 \over |q+q'|}
    \sum_{r,s=\pm} s \left[ n_B (\omega_{q'}) - n_B (\omega_{q'} + r s \omega) \right]
n_F (\epsilon_p)|_{p={m (\omega + r s \omega_{q'}) \over q+q'} - {s(q+q') \over 2}} \, . 
\end{eqnarray}
The momentum-dependent optical conductivity can easily be found using  \eqref{fund_relation}. 
We have plotted the latter for a number of different parameter sets and putting $q=0$ for definiteness.

Mainly, we are interested in the effect of different energy-momentum relations $\omega_k$ and interaction potentials $V_k$. We are looking at the conventional Fr\"ohlich model with a {longitudinal optical (LO)} phonon $\left(\omega_k,V_k\right) = \left(\Omega_0, \lambda/\left\vert k\right\vert\right)$ (for the reasons which become clear later), the acoustic phonon model $\left(\omega_k,V_k\right) = (\left\vert k\right\vert, \lambda /\sqrt{\left\vert k\right\vert})$, the small momenta BEC-polaron model \cite{Dasenbrook2012} $\left(\omega_k,V_k\right) = (c \left\vert k\right\vert, \lambda\sqrt{\left\vert k\right\vert})$  and, of course, the full BEC-polaron model as presented in the previous section.

\begin{figure}[ht]
\centering
\includegraphics[width=1\textwidth]{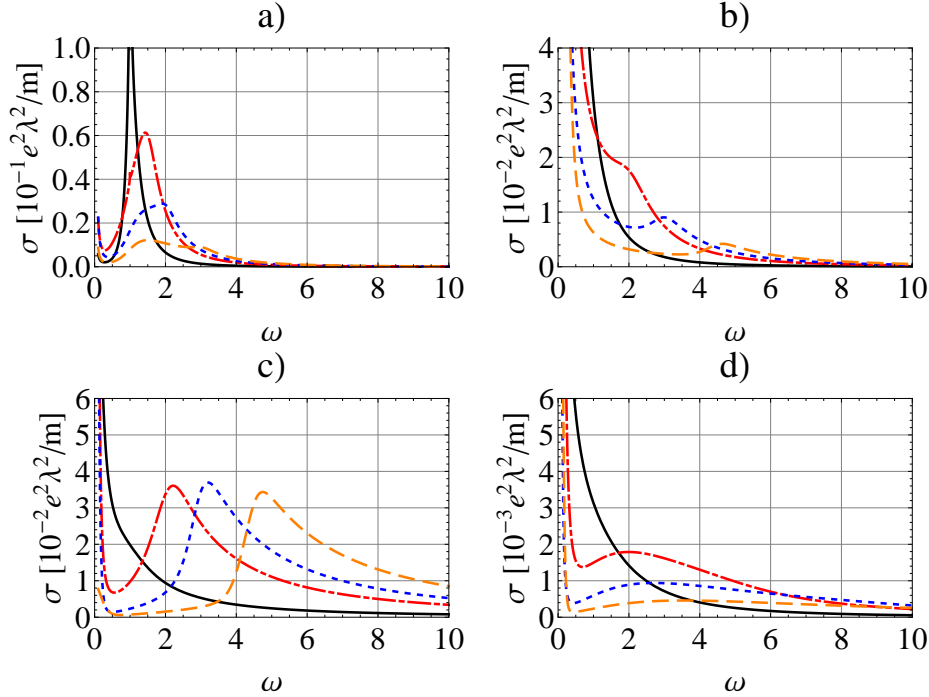}
\caption{(Color online) Real part of the conductivity for the discussed models and several chemical potentials. Panel a) the LO model ($\Omega_0=1$), b) the acoustic phonon model, c) the small momenta BEC polaron model and d) the BEC polaron model. The temperature is fixed to $\beta=10$ and the chemical potential is varied: $\mu=0,0.5,1,2$ (black solid, red dot-dashed, blue dashed and orange long-dashed line).}
\label{fig:plot1}
\end{figure}

In Fig.~\ref{fig:plot1} the conductivity is depicted for the abovementioned models for several chemical potentials.
The Fr\"ohlich model shows the well-known threshold behaviour (i.~e. $\sigma \approx 0$ for $\omega < \Omega_0$). At finite temperature, each considered model shows a divergent {behaviour} for small frequencies. This is due to the combination of the coefficient $1/\omega^3$ and the factor consisting of Bose-Einstein distribution functions. In the limit $\omega \to 0$ one usually expects to obtain the DC conductivity in the case of the `canonical' polaron system. This feature is also referred to as the Drude peak and can be traced back to the second term in \eqref{longformula_2}.
 It is the static response to a constant electric field and might indeed be rather large as compared to the rest of the spectrum. In the BEC case this Drude peak is directly related to the mobility of the impurity. 
Since the static response of {\color{black} the} polaron system is well understood, in the present study we shall concentrate on the regime $\omega>0$ of {\color{black} the} response functions.

The effect of increasing chemical potential is a shift of the peak-like structure towards higher frequencies. In the LO model an additional feature can be observed: a double peak structure emerges (one peak at $\omega \approx \Omega_0$ and another one at $\omega\approx \Omega_0 + \mu$). For {other, less trivial} energy-momentum relations $\omega_q$ this feature is washed out.  This double-peak structure has a natural explanation. The optical response measures how easy it is to excite an electron-hole pair (exciton). The existence of the first threshold is {obvious}: if there is not enough energy to overcome $\Omega_0$ then the response is suppressed. With growing $\omega$ one has to `dig' deeper into the Fermi sea. The maximal energy for the electron-hole pair is then equal to the band depth, which in the present case is equal to $\mu$.

In order to go beyond the perturbative analysis, several field theoretical approximation schemes are available. We have chosen to use the
random-phase approximation, which is good for systems with {\color{black} an} efficient screening. 
Usually this happens only in systems at high densities.
This assumption, of course, is questionable for a large fraction of experiments concerning ultracold quantum gases. For these systems (where Migdal's theorem is a priori not applicable), an approach based on vertex corrections seems to be more promising. Nonetheless, the present approximation is a good starting point for more sophisticated techniques. On the other hand, we shall see later that {\color{black} the} RPA captures many {\color{black} of the} effects found by {\color{black} the} numerically exact Monte Carlo approach. Therefore here we focus on {\color{black} the} systems with sufficiently high particle densities.

\begin{figure}[ht]
\includegraphics[width=1.\textwidth]{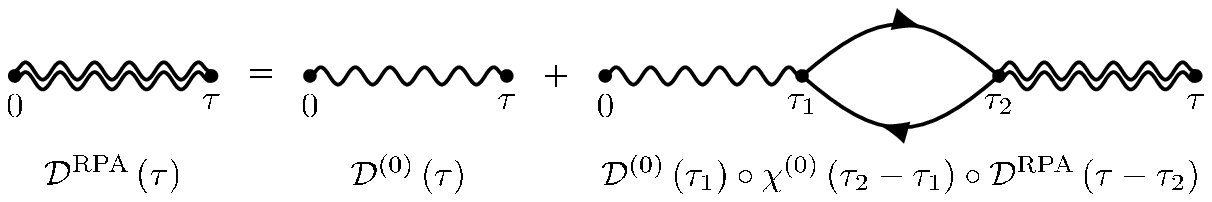}
\caption{Screened electron-phonon interaction: Dyson equation of the phonon propagator in the random phase approximation.}
\label{fig:3}
\end{figure}

The key point of the RPA approach is the replacement of the phonon Green's function by a `dressed' version, the diagrammatic expansion of which is depicted in Fig.~\ref{fig:3} and which produces  
\begin{eqnarray}
{\mathcal D}^{\mathrm{RPA}}\left(q,i\nu_n\right) = \frac{{\mathcal D}^{\left(0\right)}\left(q, i\nu_n\right)}{1-V^2_q {\mathcal D}^{\left(0\right)}\left(q,i\nu_n\right)\chi^{\left(0\right)}\left(q,i\nu_n\right)}\,.
\end{eqnarray}     
The substitution has to be done on the level of Eq.~\eqref{longformula_2}. It can be explicitly shown that this is consistent with Eq.~\eqref{longformula_1}. 
As a result we still can use Eq.~\eqref{Bragg_pert} up to the replacement $ {\cal D}^{\mathrm{(0)}}({q},i\nu_n) \to {\cal D}^{\mathrm{RPA}}({q},i\nu_n)$. Then we end up with the following expression for the Bragg spectral function: 
\begin{eqnarray}                               \label{Bragg_RPA} 
R_{\chi} (q,\omega) = & & {m \over 2 \pi |q|} n_F (\epsilon_p)|_{p={m \omega \over q} - {q \over 2}} \left[ 1 - n_F (\epsilon_p)|_{p={m \omega \over q} + {q \over 2}} \right]
\nonumber \\
& & \times \left( 1 - e^{-2 \beta m \omega} \right)
    + \frac{q^2}{\pi m \omega^4 V^2} \sum_{s=\pm} \sum_{p q'} s |V_{q'}|^2
\nonumber \\
& & \times {q'^2 \over |q+q'|} n_F (\epsilon_p) [ n_B (s \epsilon_{p+sq+sq'} - s \epsilon_p)
\nonumber \\
& & -n_B (s \epsilon_{p+sq+sq'} - s \epsilon_p - \omega) ]
\nonumber \\
& & \times \mbox{Im} D^{\mathrm{RPA},R} (q', s \epsilon_{p+sq+sq'} - s \epsilon_p - \omega) \, . 
\end{eqnarray}

\begin{figure}[th]
\includegraphics[width=1.\textwidth]{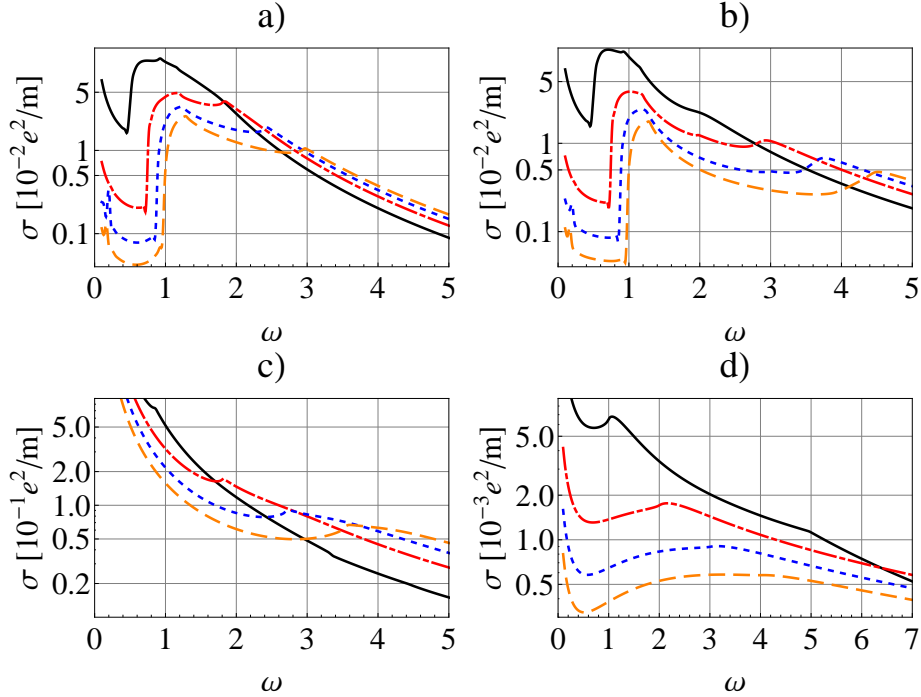}
\caption{(Color online) Optical conductivity in the RPA approximation for fixed interaction strength $\lambda=1$ and varying chemical potential $\mu=0.5,1,1.5,2$ (black solid, red dot-dashed, blue dashed and orange long-dashed line). The order of the graphs is the same as in the figures above.}
\label{plot:3}
\end{figure}

\begin{figure}[th]
\includegraphics[width=1\textwidth]{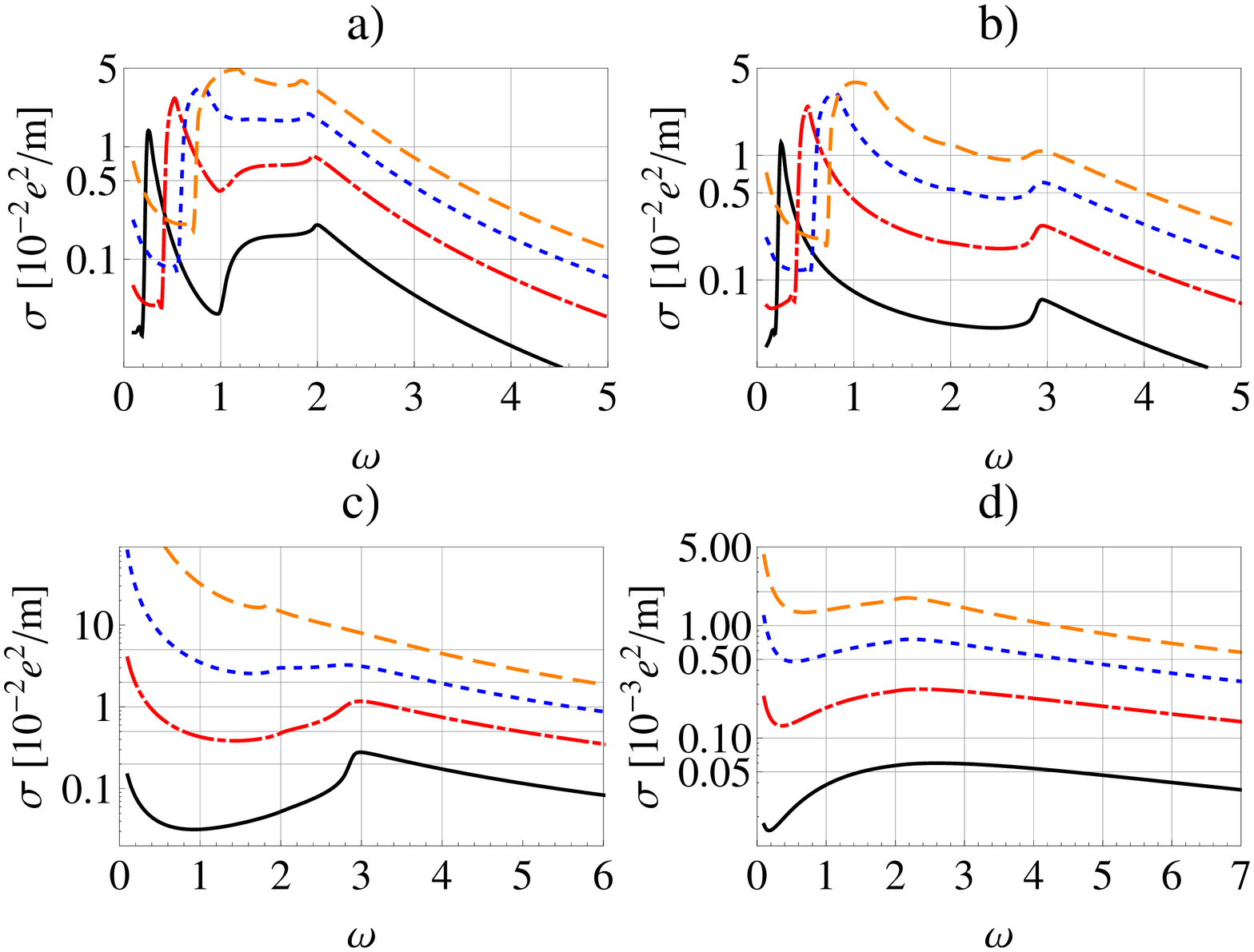}
\caption{(Color online) Optical conductivity in the RPA approximation for fixed chemical potential $\mu=1$ and varying interaction strength $\lambda=0.25,0.5,0.75,1$ (black solid, red dot-dashed, blue dashed and orange long-dashed line). The order of the graphs is the same as in the figures above.}
\label{plot:4}
\end{figure}

Figs.~\ref{plot:3} and \ref{plot:4}  show the conductivity for several parameter constellations. 
As already seen in the perturbative calculation, Fig.~\ref{plot:3} shows a pronounced shift of the secondary peak with changing chemical potential. The most prominent features, however, are seen for changing coupling strength, see Fig.~\ref{plot:4}. In the case of the LO phonon the positions of the maxima in the two-peak structure are consistent with the perturbative calculation and are located at $\omega=\Omega_0$ and $\omega = \Omega_0 + \mu$. In the case with {\color{black} a} linear phonon dispersion relation -- for the acoustical phonons and a BEC system with linearized dispersion [panels b) and c)] the position of the secondary peak is different. In fact its location is still very sensitive to changing $\mu$ but without any perceivable shift at different $\lambda$. 

Another interesting feature is the  strong suppression of the spectral weight {at intermediate frequencies}, which we would like to call \emph{pseudogap}. It  is seen as the first minimum in the spectrum, which develops in the non-BEC case. As is clearly seen in Fig.~\ref{plot:3} a) and b) it shows itself as a frequency threshold similar to the one seen in the above LO case (that is why we consider that situation along with the different BEC models). The pseudogap width grows with {\color{black} the} coupling strength and is directly connected to the energy stored in the fully developed polaron state which has to be destroyed prior to excitation of the impurity.

In the BEC case,  no pronounced pseudogap formation is observed, there is just a small dip in the spectral weight. From the mathematical point of view the
reason for the absence of {\color{black} the} pseudogap is the different $k$-dependence of coupling constant $V_k$. In the LO and acoustic phonon models, $V_k$ diverges at $k \rightarrow 0$, which is advantageous for creating a gap,
while for the small momenta and ordinary BEC polaron models, $V_k$ vanishes at $k \rightarrow 0$. This means {\color{black} that} a much stronger coupling is needed to open a gap.

In contrast to the gap development, the Drude peak tends to be enhanced.
This is consistent with the outlined picture since according to the Drude contribution shown e.~g. in Eq.~\eqref{Bragg_RPA} it is proportional to the impurity mass, which is enhanced as soon as the polaron starts to form. 

\begin{figure}[th]
\includegraphics[width=1.\textwidth]{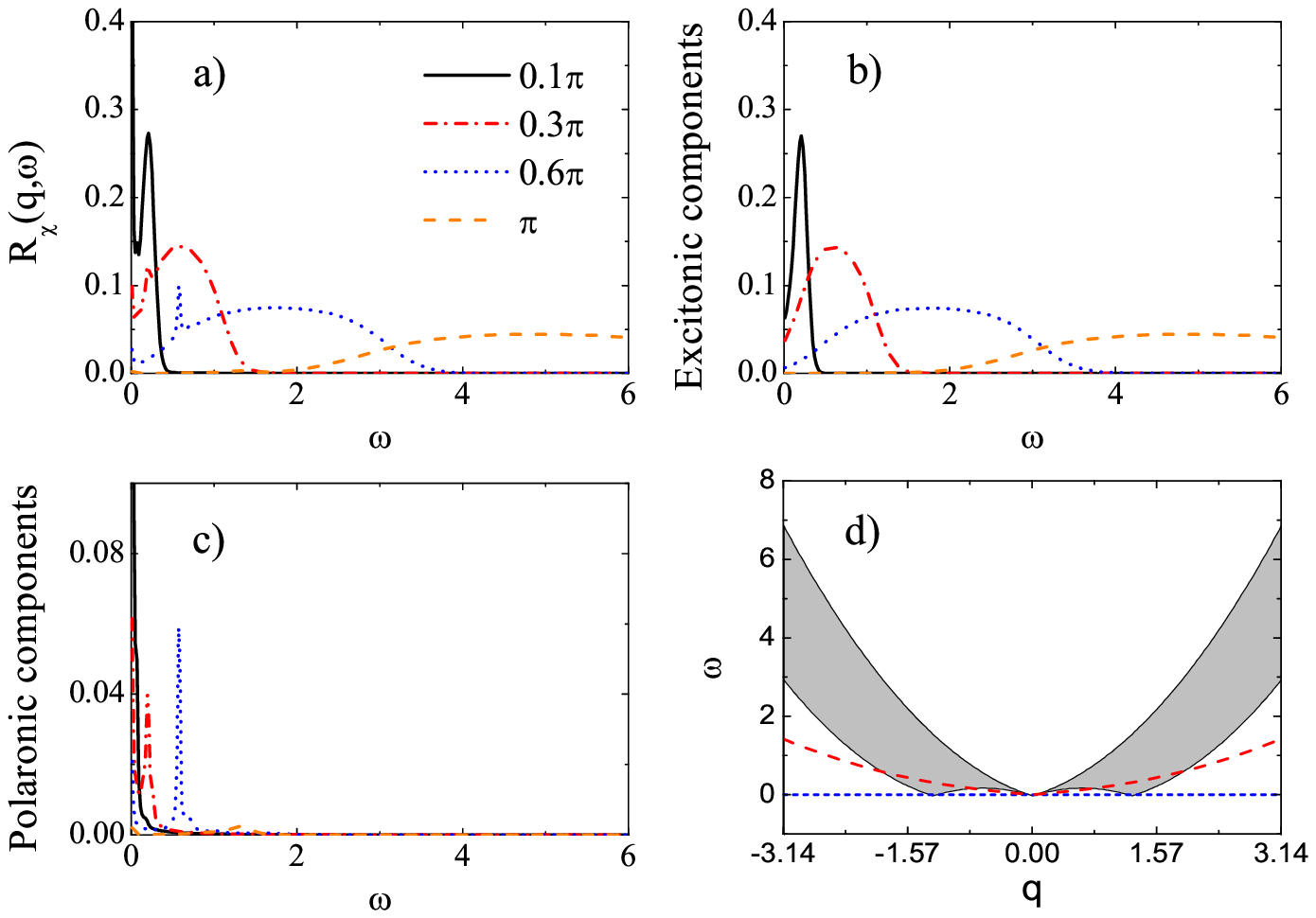}
\caption{
(Color online) Characteristic structures in the Bragg spectra of a boson-fermion mixture in {\color{black} a} 1D BEC polaron model, calculated by {\color{black} the} RPA at $\beta$=10 and $\lambda$=0.08.
The Fermi momentum is at 0.2$\pi$, corresponding to $\mu$=0.197.
Panel a) shows the Bragg spectral functions for various wave vectors.
Panels b) and c) reveal respectively the excitonic and polaronic components in a).
Panel d) depicts the Drude (blue dots), Bogolyubov (red dashed) and excitonic (shaded) band dispersions in the Bragg spectra.}
\label{Fig_RPA_1}
\end{figure}

Now we turn to the Bragg spectra at $q\neq 0$. For not too strong boson-fermion coupling it is plotted in Fig.~\ref{Fig_RPA_1}. 
As can be seen already at Eq.\eqref{Bragg_pert} the spectrum has two contributions. One is due to {\color{black} the} excitation of electron-hole pairs (excitons), which is proportional to $n_F(\epsilon_{p})[1-n_F(\epsilon_{p+q})]$ and independent on the coupling strength. On the contrary, the other part is interaction-dependent and can, in turn, be subdivided into the Drude peak at $\omega=0$ as well as a phonon peak which follows the Bogolyubov dispersion, see Fig.~\ref{Fig_RPA_1} c). All these features become more lucid if one plots the dispersion relations of all elementary excitations in the system, see Fig.~\ref{Fig_RPA_1} d). The shaded region for $\omega$ between $\left| |q| \sqrt{2 \mu / m} - q^2 / (2 m) \right|$ and $|q| \sqrt{2 \mu / m} + q^2 / (2 m)$ matches perfectly with the wide excitonic plateau.

\section{Monte Carlo simulations}
\label{sec:MC}

As we have seen in the previous section, the optical spectra change considerably in the case of strongly interacting systems. In order to corroborate the RPA results and clarify the physical picture one has to employ more advanced techniques.
Quantum Monte Carlo simulation method is one of such powerful approaches which enable to access the dynamical properties of the system \cite{DeFilippis2012a,Mishchenko2003a}.
Although in some cases it is subjected to limitations such as finite size and the sign problem, this method has been successfully used in many polaron related problems \cite{Mishchenko2004a,Hohenadler2012a}.
Its results have also been considered as benchmarks {in regimes} where exact solutions are not available.

Our QMC simulation is based on a path integral formulation of the dynamical correlation functions.
We shall follow the theoretical treatment worked out in a previous work \cite{Ji2004} with an extension of it to the conjugate momentum space which is convenient in our present study.

\subsection{Formulation of the path integral}

For our purposes it is very convenient to reduce the phonon field to the original set of harmonic oscillator operators as 
$q_k = 1 / \sqrt{2 m_p \omega_k} (b_{-k}^{\dag} + b_k)$, and
$p_k = i \sqrt{m_p \omega_k / 2} (b_{-k}^{\dag} - b_k)$,
to replace the phonon creation and annihilation operators in Hamiltonian (1), with $m_p$ being the effective oscillator mass.
So that the Fr\"ohlich Hamiltonian is rewritten as
\begin{eqnarray}
H &=& \sum_q \left(E_q - \mu \right) a_q^{\dag} a_q
+ \sum_k \left( {p_k^2 \over 2 m_{p}} + {1 \over 2} m_p \omega_k^2 q_k^2 \right)
\nonumber \\
&+& \sum_q \sum_{k \neq 0} V_k \sqrt{m_p \omega_k \over 2}
\left( a_{q+k}^{\dag} a_q q_k + \mbox{H.c.} \right).
\end{eqnarray}
By using the standard Trotter's decoupling scheme, we represent the Boltzmann operator in a path integral form,
\begin{eqnarray}
e^{- \beta H} &\rightarrow&
    \int \mathcal{D} x T_{\tau} \exp \left \{ - \int_0^{\beta} d \tau [ h_e (\tau, x) + h_{ph} (\tau, x) ] \right \}
    \nonumber \\
& \times & \prod_k | x_k (\beta) \rangle \langle x_k  (0)| ,
\end{eqnarray}
where
\begin{eqnarray}
h_e (\tau, x) &=& \sum_q \left(E_q - \mu \right) a_q^{\dag} (\tau) a_q (\tau)
    \nonumber \\
& + & \sum_q \sum_{k \neq 0} \sqrt{m_p \omega_k \over 2}
\left[ a_{q+k}^{\dag} (\tau) a_q (\tau) q_k (\tau) + \mbox{H.c.} \right] \, ,
    \nonumber \\
h_p (\tau, x) &=& \sum_k \left \{ {m_p \over 2} \left[ {\partial x_k (\tau) \over \partial \tau} \right]^2
    + {1 \over 2} m_p \omega_k^2 x_k^2 (\tau) \right \} .
\end{eqnarray}
Here $\tau$ is the imaginary time, $| x_k \rangle$ is the eigenstate of the operator $q_k$ with eigenvalue  $x_k$ satisfying the eigen-equation $q_k | x_k \rangle = x_k | x_k \rangle$.
Then we define the time evolution operator $U_x (\tau)$ along {\color{black} the} path $x$ as
\begin{eqnarray}
U_x (\tau) = T_{\tau} \exp \left \{ - \int_0^{\tau} d \tau' h_e [\tau', x(\tau') ] \right \} .
\end{eqnarray}
On this path, the free energy ($\equiv F_x$) is evaluated by a trace over the fermionic part of the Boltzmann operator,
\begin{eqnarray}
e^{- \beta F_x} &=& e^{- \int_0^{\beta} d \tau h_p (x)} \mbox{Tr} [U_x (\beta)]
\nonumber \\
&=& e^{- \int_0^{\beta} d \tau h_p (x)} \det |{\bf I + U}_x (\beta)| ,
\end{eqnarray}
where ${\bf U}_x (\tau)$ is the matrix representation of the time evolution operator $U_x (\tau)$.
The partition function ($\equiv Z$) and total free energy is obtained by integrating out the bosonic field {along the path} $x$,
\begin{eqnarray}
Z = e^{- \beta F} = \int \mathcal{D} x e^{- \beta F_x} .
\end{eqnarray}
The expectation value of an operator $O$ is given by
\begin{eqnarray}
\langle O \rangle &=& {1 \over Z} \int \mathcal{D} x e^{- \beta F_x}
    \langle O \rangle_x ,
\end{eqnarray}
where $\langle O \rangle_x$ is the average along a path $x$,
\begin{eqnarray}
\langle O \rangle_x = \frac{\mbox{Tr} [U_x (\beta) O]}{\mbox{Tr} [U_x (\beta)]} .
\end{eqnarray}

In this notation, the path integral form for the single particle Matsubara Green's function is given by:
\begin{eqnarray}
G (q, \tau; q', \tau') &=& {1 \over Z} \int \mathcal{D} x e^{- \beta F_x}
    G_x (q, \tau; q', \tau') ,
\nonumber \\
G_x (q, \tau; q', \tau') &=& - \langle T_{\tau} \hat{a}_q (\tau) \hat{a}_q'^{\dag} (\tau') \rangle_x ,
\end{eqnarray}
where $\hat{a}_q (\tau)$ is the Heisenberg representation of $a_q$, defined {as usual} by
\begin{eqnarray}
\hat{a}_q (\tau) = U_x^{\dag} (\tau) a_q U_x (\tau) .
\end{eqnarray}
By solving the equation of motion for the involved operators \cite{Tomita1997}, we get the path-dependent Green's function ($\beta \ge \tau \ge \tau' \ge 0$),
\begin{eqnarray}
G_x (q, \tau; q', \tau') =
    - \left\{ {\bf U}_x (\tau) [1 + {\bf U}_x (\beta)]^{-1} \right.
\left. {\bf U}_x^{-1} (\tau') \right\}_{q,q'} .
\end{eqnarray}
The optical conductivity and absorption spectrum are derived from the current-current correlation function \eqref{momdepcond}, which can be rewritten as
\begin{eqnarray}
&& \Pi (q, \tau; q', \tau') = {1 \over Z} \int \mathcal{D} x e^{- \beta F_x}
    \Pi_x (q, \tau; q', \tau') ,
    \nonumber \\
&& \Pi_x (q, \tau; q', \tau')
    = - {1 \over V} \langle T_\tau \hat{j}^{\dag} (q, \tau) \hat{j} (q', \tau') \rangle_x
    \nonumber \\
&& = - {e^2 \over m_e^2 V} \sum_{kk'}
\left( k + {q \over 2} \right) \left( k' + {q \over 2} \right)
    \nonumber \\
&& \times \left[ G_x (k', \tau'+\beta; k, \tau) G_x (k+q, \tau; k'+q, \tau') \right.
    \nonumber \\
&& + \left. G_x (k, \beta; k+q, 0) G_x (k'+q, \beta; k', 0) \right] .
\end{eqnarray}
In the last line, the time-dependent Bloch-De Dominicis theorem \cite{Tomita1997} has been employed to decouple the many-body operators into a product of bilinear components.
In the analogous manner, the Bragg spectrum is related to the density-density correlation function \eqref{autodensity}, which can be expressed in the form like
\begin{eqnarray}
&& \chi (q, \tau; q', \tau') = {1 \over Z} \int \mathcal{D} x e^{- \beta F_x}
    \chi_x (q, \tau; q', \tau') ,
    \nonumber \\
&& \chi_x (q, \tau; q', \tau')
    = - {1 \over V} \langle T_\tau \hat{\rho}^{\dag} (q, \tau) \hat{\rho} (q', \tau') \rangle_x
    \nonumber \\
&& = - {1 \over V} \sum_{kk'}
    \left[ G_x (k', \tau'+\beta; k, \tau) G_x (k+q, \tau; k'+q, \tau') \right.
    \nonumber \\
&& + \left. G_x (k, \beta; k+q, 0) G_x (k'+q, \beta; k', 0) \right] .
\end{eqnarray}

\subsection{Numerical results on dynamical responses}

In our numerical calculation, the path integral is performed by the QMC simulation method along with the matrix factorization and QDR decompositions in quad precision to reduce the numerical errors.
During the data acquisition, the dynamical correlation functions $\Pi_x (q, \tau)$ and $\chi_x (q, \tau)$ are measured after every 100 updates.
We {gather} a total {number of about} 5000-20000 samples until the simulation converges.
In addition, extra 100-500 samples are swept to thermalize the system from a starting configuration to the equilibrium regime.
From the QMC data of current autocorrelation functions, we derive the absorption spectrum $R_{\Pi} (q, \omega)$ {solving} the integral equation
\begin{eqnarray}
\label{eq:mc1}
\Pi (q, \tau) = - \int_{- \infty}^{\infty} d \omega R_{\Pi} (q, \omega)
\frac{e^{- \tau \omega}}{1 - e^{- \beta \omega}} 
\end{eqnarray}
{\color{black} by inversion.}
Then the optical conductivity is easily obtained from
\begin{eqnarray}
\mbox{Re} [\sigma (\omega)] = \lim_{q \rightarrow 0} {\pi \over \omega} R_{\Pi} (q, \omega) .
\end{eqnarray}
Similar to Eq.~\eqref{eq:mc1}, the Bragg spectral function $R_{\chi} (q, \omega)$ is extracted from the density autocorrelation $\chi_x (q, \tau)$ via
\begin{eqnarray}
\label{eq:mc2}
\chi (q, \tau) = - \int_{- \infty}^{\infty} d \omega R_{\chi} (q, \omega)
\frac{e^{- \tau \omega}}{1 - e^{- \beta \omega}} .
\end{eqnarray}
In order to solve the integral equations Eq.~(\ref{eq:mc1}) and Eq.~(\ref{eq:mc2}) for the two-particle correlation functions, we have developed a renormalizing iterative fitting method.
Some details of this algorithm are described in \ref{sec:app_ac}. Moreover, we have performed simulations for different system sizes and did not observe perceivable finite size effects for systems with more than 20 sites.

\begin{figure}[htbp]
\centering
\includegraphics{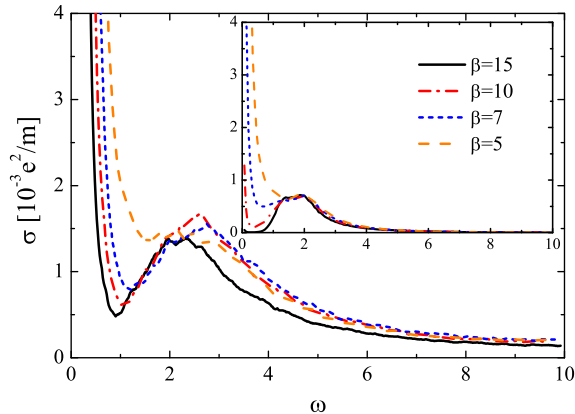}
\caption{
(Color online) Temperature dependence of {\color{black} the} real part of {\color{black} the} optical conductivity for {\color{black} the} LO model with 12 particles on a 1D lattice of 25 sites.
The coupling constant is fixed to $\lambda$=0.1 and $\Omega_0=1$.
The main graph presents the spectra from QMC simulation, whereas the inset shows the RPA results.}
\label{fig:MC1}
\end{figure}

In the first step we would like to compare the QMC simulation results with those of the RPA.
In Fig.~\ref{fig:MC1} we plot the optical conductivity in the case of weak coupling and different temperatures. 
The RPA calculation is conducted on an infinitely large system, while the QMC simulation is performed on a system of 12 particles on 25 sites with open boundary conditions.
In both cases, the Fermi momentum is about $\pi/2$.
Since QMC can only deal with models of finite size, we have performed simulations with momenta confined to the region $-\pi \le q \le \pi$.
In Fig.~\ref{fig:MC1}, in both QMC (main plot) and RPA (inset) results one immediately recognizes the two features: a Drude peak nearby $\omega$=0, and a phonon peak at $\omega \gtrsim \Omega_0$. Although there is no perfect numerical match between the results
-- for instance according to the QMC data the Drude peak lies at $\omega \neq 0$ (this is a finite size effect), both 
methods show the same qualitative behaviour. E.~g.
the Drude peak declines with the decreasing temperature.

\begin{figure}[htbp]
\centering
\includegraphics{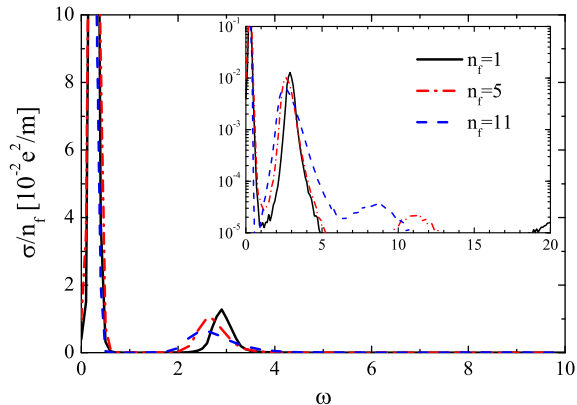}
\caption{
(Color online) Real part of {\color{black} the} optical conductivity of a 1D LO model with 25 polaronic states at different filling levels, when $\lambda$=0.5, $\beta$=10 and $\Omega_0=1$.
The three spectra in the main panel correspond to filling fractions: single particle (black solid), 5 (red dot-dashed) and 11 (blue dashed) particles.
The inset displays the same results in semi-logarithmic scale to expose the fine structures.}
\label{fig:MC2}
\end{figure}

In Fig.~\ref{fig:MC2}, we study the effect of doping on the absorption spectrum.
The total number of fermions ($\equiv n_f$) is adjusted to 1, 5 and 11 by changing the chemical potential.
It follows from the $f$-sum rule of optical conductivity \cite{Tempere2001a,Tempere2001b}, that the spectral weight is modulated by the charge density.
In order to factor out this effect, in Fig.~\ref{fig:MC2} we renormalize the spectra by $n_f$.

The main graph shows the spectra in linear scale, and the inset in semi-logarithmic scale so as to resolve some fine structures in the spectra.
Here one can clearly distinguish a small island apart from the Drude (near $\omega$=0) and phonon peaks (at $\omega$=2.5$\sim$3.0).
This island is due to the high-order electron-phonon scattering processes and is located at the high energy {\color{black} tail of the spectrum}.
When we increase $n_f$, we actually introduce more polarons into the system.
So that the many-body effect gradually emerges as the polarons begin to interact with each other.
As a consequence, the single-phonon peak is somewhat broadened with increasing $n_f$, and the multi-phonon island is shifted towards smaller energies with a growing amplitude.
This tendency suggests that the polaron effect can be reinforced by increasing the density of fermions.

\begin{figure}[htbp]
\centering
\includegraphics{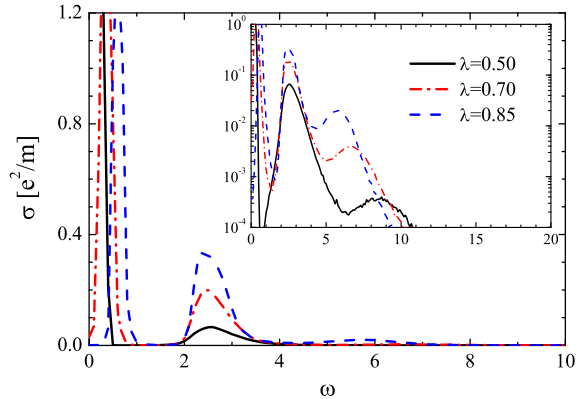}
\caption{
(Color online) Real part of {\color{black} the} optical conductivity of a 1D LO model with 11 impurities on a lattice of 25 sites at $\beta$=10 and $\Omega_0$=1.
The three spectra in the main frame correspond to different coupling strengths $\lambda$: 0.5 (black solid), 0.7 (red dot-dashed) and 0.85 (blue dashed).
The inset displays the same results in semi-logarithmic scale to expose the {\color{black} details of the spectra}.}
\label{fig:MC3}
\end{figure}

Next we investigate the dependence on the interaction strength, varying $\lambda$ between 0.25 and 0.85. The last value is of the order of the band width and thus we expect it to drive the system deep into the strong coupling regime.
Fig.~\ref{fig:MC3} shows the results for the LO phonon situation, again on a lattice with 25 sites and 11 impurities.
All curves share the general feature of a peak around $\omega \sim 2.3$, which is just the usual threshold frequency given by the phonon frequency plus the level spacing due to the finite size. This compares well with the RPA calculation. The secondary peaks due to excitation of impurities which lie deeper in the Fermi sea are much less pronounced, but undergo a shift towards lower frequencies for growing $\lambda$. This is much better seen in the inset of the figure.

More fundamental differences can be observed for the Drude peak. Starting with $\lambda=0.5$ it is moving towards finite energies while the spectral weight at $\omega=0$ almost completely vanishes. Thus a pseudogap opens up indicating that the system becomes `insulating'. This effect is fully covered by the RPA calculation and reflects the polaron binding energy.

\begin{figure}[htbp]
\centering
\includegraphics{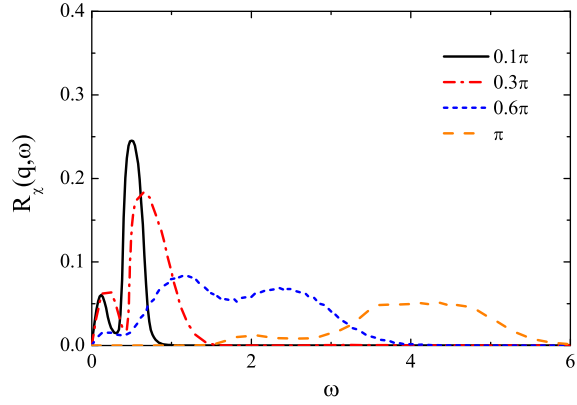}
\caption{
(Color online) Bragg spectra at different wave vectors for a 1D BEC polaron model involving 4 fermions on 21 sites, computed by a QMC simulation at $\lambda$=0.08, $\beta$=10.
Comparable RPA results for the same parameters are presented in Fig.~\ref{Fig_RPA_1}.}
\label{fig:MC4}
\end{figure}

Now we turn to the case of the BEC with the dispersion relation Eq.~\eqref{phonondisp}. 
Fig.~\ref{fig:MC4} shows the Bragg spectra simulated for the parameter constellation used in Fig.~\ref{Fig_RPA_1}.
Comparing Fig.~\ref{fig:MC4} with Fig.~\ref{Fig_RPA_1}(a), one can easily recognize all  essential components of Bragg spectra, i.~e. the Drude peak, phonon excitation and exciton plateau are well captured.
Similarly to the optical conductivity, the Drude peak is slightly shifted in QMC results  due to the finite energy level spacing.

Another feature of the QMC data, namely the vanishing spectral weight towards higher frequencies (in our case $\omega {\color{black} \gtrsim 5}$) is a quite natural consequence of the cut-off procedure used in the numerics -- we have set $\omega_c=\pi$. Thus, the only genuine difference between the QMC and RPA results is the enhanced spectral weight due to phonons. This is an artefact of RPA and this discrepancy vanishes for weaker coupling.

\begin{figure}[htbp]
\centering
\includegraphics{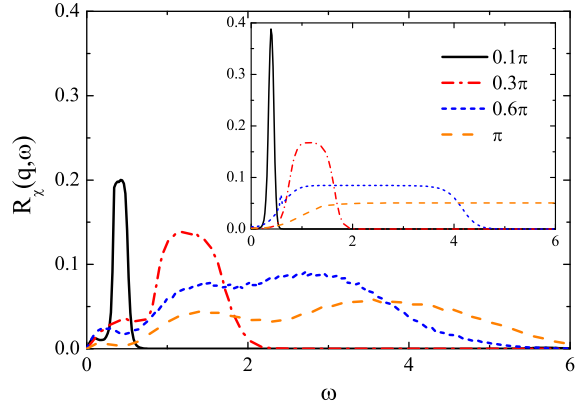}
\caption{
(Color online) Bragg spectra at different wave vectors for a 1D BEC polaron model involving 8 fermions on 21 sites, when $\lambda$=0.08, $\beta$=10.
The spectra in the main frame are computed by a QMC simulation, and the inset by an RPA calculation.}
\label{fig:MC5}
\end{figure}

In Fig.~\ref{fig:MC5}, we increase the total number of fermions to 8 within the 21-site system.
We keep $\lambda$=0.08 and $\beta$=10, the same as in Fig.~\ref{fig:MC4}.
The main graph displays the Bragg spectra from QMC, and the inset from RPA for comparison.
As already shown above, the excitonic component of Bragg spectrum is located in the region
$\left| |q| \sqrt{2 \mu / m} - q^2 / (2 m) \right| \le \omega \le |q| \sqrt{2 \mu / m} + q^2 / (2 m)$.
Its spacial range varies with fermion density, making the plateau distinguishable from the polaron peaks which do not shift with $\mu$.
In Fig.~\ref{fig:MC5}, these properties are consistently reproduced by both RPA and QMC calculations.

\begin{figure}[htbp]
\centering
\includegraphics{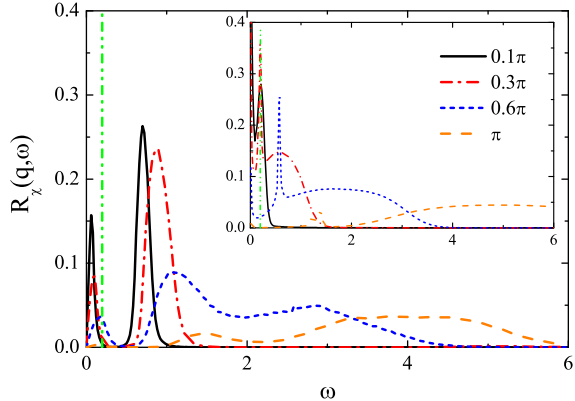}
\caption{
(Color online) Bragg spectra at different wave vectors for a 1D BEC polaron model involving 4 fermions on 21 sites, when $\lambda$=0.16, $\beta$=10.
The spectra in the main graph are computed by a QMC simulation and the inset shows RPA results.
The vertical green dot-dot-dashed line labels the position of Fermi energy.}
\label{fig:MC6}
\end{figure}

For growing interaction strength the predicting power of RPA rapidly degrades as can be seen in Fig.~\ref{fig:MC6}.
It fails to track the frequency renormalization and the widening of the phonon peak {\color{black} is}  clearly seen in the QMC data. Interestingly, for increasing coupling a formation of a pseudogap at small $\omega$ can clearly be seen {\color{black} as well}.
 On the mathematical level this pseudogap can be understood as a `peak repulsion' of the Drude and phonon contributions, which overlap around $\omega\approx 0$ in the non-interacting case and are subject to avoided crossing as soon as the boson-fermion interaction is switched on.
While {\color{black} the} RPA results of previous section Eq.~\eqref{Bragg_RPA} (plotted in inset) point towards weak spectral weight suppression for small $q$  and a larger one for finite momenta,
QMC data (main graph) suggest that this gap is in reality much stronger with almost vanishing spectral weight.

\begin{figure}[htbp]
\centering
\includegraphics{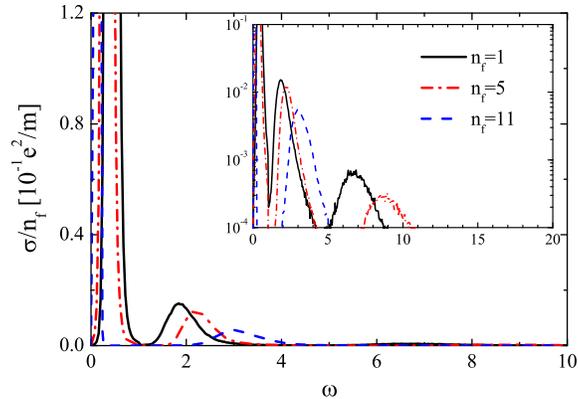}
\caption{(Color online) Real part of {\color{black} the} optical conductivity for a 1D system
involving fermions and bosonic excitations in BEC, when $\lambda$=0.25, $\beta$=10.
The three spectra in the main frame correspond to different doping levels: single (black solid), 5 (red dot-dashed) and 11 (blue dashed) fermonic impurities.
The inset displays the same result in a semi-logarithmic representation.}
\label{fig:MC7}
\end{figure}

As can be seen in Eq.~\eqref{fund_relation}, the Bragg spectral function becomes ill-defined at $q \rightarrow 0$.
In this case the absorption spectral function turns out to be more suitable for the investigation.
Although the absorption spectrum might not be directly measurable in a cold atomic setup, from the theoretical point of view it provides complementary information on the dynamical properties which are inaccessible for the Bragg spectroscopy.
In the rest of this section, we focus on the optical conductivity of the BEC polaron model with {\color{black} a special} attention to the formation of pseudogap.
In Fig.~\ref{fig:MC7}, the real part of optical conductivity of a 1D 21-site system is depicted for  filling fractions $n_f{\color{black}/N}$=1/21, 5/21 and 11/21, respectively.
Here the coupling constant is set to $\lambda$=0.25, and the inset displays the spectra in a semi-logarithmic scale so as to amplify the fine structures.
From the black curve with $n_f$=1, one can infer that $\lambda$=0.25 is already of intermediate strength because there is a clear pseudogap at $\omega$=0. This is different from the RPA calculation. Thus we conclude that for intermediate to large coupling strengths RPA overestimates the screening effects, leading to the pseudogap formation.

With $n_f$ increasing to 5, one finds that the gap width and depth change, as shown by the dot-dashed (red) curves.
If we go on increasing $n_f$ to 11, as illustrated by the dashed (blue) curve, the gap eventually closes.
This can be explained in the following way. 
The gap reflects the energetic cost required to set free a fermion from its bounded polaron state, which is higher for stronger boson-fermion interactions. Those tend to be screened for growing fermion densities though, with decreasing polaron binding energy as a result.
Such evolution of the gap cannot be observed in the RPA calculation.  On the one hand, it is done for the continuum model from the outset and so there is no finite energy level spacing in the first place. On the other hand RPA, being a high-density approximation, is known to be able to perfectly describe the screening effects so that it is quite natural that there the pseudogap is strongly suppressed.
As the state-of-the-art apparatus allows for generation of rather short optical lattices we believe that the pseudogap could be observed experimentally.

Finally, there is yet another {\color{black} possible} explanation for the pseudogap formation{\color{black}, which we can rule out though}. 
The phonon modes of momenta $k = \pm 2q_{\rm F}$ {\color{black} might be} responsible for a gap opening due to the Peierls instability, where $q_{\rm F}$ is the Fermi momentum.
However, a precise estimation of the corresponding gap shows that it cannot come from the Peierls transition.
The transition temperature is lower than the temperature regime here, and hence our system is basically located in a gapless phase.
To make sure we are away from the Peierls instability, we {\color{black} have} checked the one-fermion spectral function (not shown here), which can probe the gap in fermion energy band.
We have found that a band gap of this type cannot be generated.

\begin{figure}[htbp]
\centering
\includegraphics{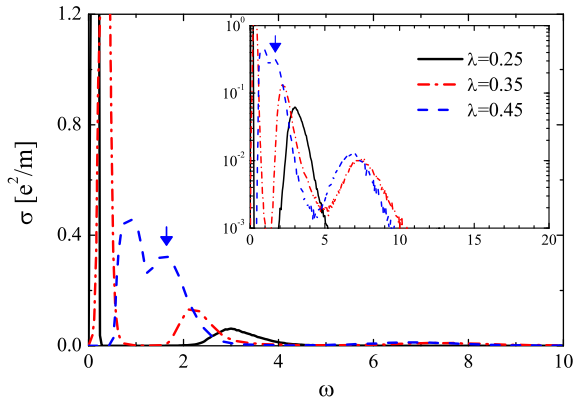}
\caption{(Color online) Real part of {\color{black} the} optical conductivity for a 1D model with 11 fermionic impurities in a system of 21 sites at $\beta$=10.
The three spectra in the main frame correspond to different coupling strength $\lambda$: 0.25 (black solid), 0.35 (red dot-dashed) and 0.45 (blue dashed).
The inset shows the same result in semi-logarithmic representation.
Blue arrows in the figure indicate the phonon shoulders.
}
\label{fig:MC8}
\end{figure}

The difficulty to open a gap at high $n_f$ in Fig.~\ref{fig:MC7} can be partially eliminated by increasing the coupling strength $\lambda$.
The attraction between fermions and bosons produces a negative coupling energy and tends to compensate the energy gain from phonon creation processes.
In the 1D system, this coupling would end up with a gap opening, provided that $\lambda$ is large enough to stabilize the phonon modes.
In Fig.~\ref{fig:MC8}, we examine this property from a view of optical conductivity for the BEC polarons.
We again consider 11 fermions immersed in a 1D trap of BEC with 21 sites.
The $\lambda$-dependence is similar to that of {\color{black} the} LO model in Fig.~\ref{fig:MC3}.
Here $\lambda$=0.25 (black solid), 0.35 (red dot-dashed) and 0.45 (blue dashed) correspond to weak, intermediate and strong couplings, respectively.
However, in Fig.~\ref{fig:MC5}, the phonon peak is highly modified not only in its shape but also in its position.
When $\lambda$=0.45, the phonon peak even plunges into the Drude peak, giving rise to a broad shoulder (pointed out by arrows (blue) in the figure).
Such a strong delocalization effect on the phonon peak is absent in the conventional LO Fr\"ohlich model.
It can be attributed to the $k$-dependence of coupling $V_k$ in the BEC polaron model.
The most important features of the spectra are concentrated around a few $k$ values.
Since $V_k= \lambda [(\xi k)^2 / ((\xi k)^2 + 2)]^{1/4}$, if $\lambda$ increases to a larger value $\lambda'$, we can find a smaller $k$ to keep $V_k$ invariant, i.e. $V_{k'} (\lambda') = V_k (\lambda)$,  thus applying a `discrete' version of renormalization transformation.
That means we can attain roughly the same coupling energy by enhancing $\lambda$ and {\color{black} simultaneously} reducing $k$.
Therefore, the phonon excitations gradually concentrate to the small $k$ regime, and the low energy phonon modes become more favorable as $\lambda$ increases.
This displacement of phonon peak leads us to conclude that for large coupling, those phonon modes of small momenta or long wavelengths play more important role in the dynamical response of the system.

\section{Discussion and conclusions}
\label{sec:discussion}

We have analyzed the spectrum of the correlation function of particle currents in a number of interacting mixtures of bosons and fermions. We {\color{black} have} considered different kinds of couplings and dispersion relations for the constituent subsystems, which model {\color{black} the} electrons in semiconductors coupled to LO phonons, acoustical phonons as well as fermionic impurities which are immersed into a BEC and which interact with Bogolyubov modes of the condensate.  While for the solid state realizations the quantity we calculate is the optical response, in the BEC case the correlation function of currents gives a direct access to the Bragg spectra of impurities. 

While in the weak coupling case we recover all of the known physics, we find distinct effects in the situations with intermediate to strong interactions. Especially a pseudogap (with respect to particle-hole pair excitation) formation could be found using QMC simulations in 1D. We speculate that this effect is due to the finite energy stored in the polaron state, which is released/absorbed during the fermion excitation process. 

Our approaches allow for an extension to a number of realistic experimental setups. We expect that the effects we predict could, for example, be investigated in binary systems which are similar to those used in Refs.~\cite{PhysRevLett.102.230402} or \cite{PhysRevA.85.042721}.

The authors thank Wim Casteels, Sergei Klimin, Jozef Devreese, Tobias Schuster, Raphael Scelle and Markus Oberthaler for many inspiring discussions. Financial support was provided by the DFG under Grant No. KO 2235/5-1 and by the `Enable Fund', the CQD and the HGSFP of the University of Heidelberg.

\appendix

\section{Relation between optical absorption and Bragg spectra}
\label{sec:app_spec}

We start with the definition of density-density correlation function,
\begin{eqnarray} \label{eq:chi1}
\chi ({\bf q}, i \omega_n) = - {1 \over V} \int_0^{\beta} d \tau e^{i \omega_n \tau}
\langle T_\tau \rho^{\dag} ({\bf q}, \tau) \rho ({\bf q}, 0) \rangle .
\end{eqnarray}
Integrating Eq.~\eqref{eq:chi1} by part on variable $\tau$, 
\begin{eqnarray} \label{eq:parint1}
\int u dv = uv - \int v du ,
\nonumber \\
u = \rho^{\dag} ({\bf q}, \tau) ,
v = \frac{e^{i \omega_n \tau}}{i \omega_n} ,
\end{eqnarray}
we get
\begin{eqnarray} \label{eq:chi2}
\chi ({\bf q}, i \omega_n) &=& - {1 \over i \omega_n V} \langle [\rho ({\bf q}), \rho^{\dag} ({\bf q})] \rangle
    + {1 \over i \omega_n V} \int_0^{\beta} d \tau e^{i \omega_n \tau}
\nonumber \\
& & \times \langle T_{\tau} \frac{d \rho^{\dag} ({\bf q}, \tau)}{d \tau} \rho ({\bf q},0) \rangle .
\end{eqnarray}
The commutator in the first term gives zero
while the correlator in the second term depends only on the difference of $\tau$, hence can be rewritten into
\begin{eqnarray} \label{eq:parint2}
\langle T_{\tau} \frac{d \rho^{\dag} ({\bf q}, \tau)}{d \tau} \rho ({\bf q},0) \rangle
= \langle T_{\tau} \left[ \frac{d \rho^{\dag} ({\bf q}, \tau')}{d \tau'} \right]_{\tau'=0} \rho ({\bf q}, -\tau) \rangle .
\nonumber \\
\end{eqnarray}
So that in the term $\rho ({\bf q}, -\tau)$ the integration by parts can be done again yielding
\begin{eqnarray} \label{eq:chi3}
\chi ({\bf q}, i \omega_n) &=& {1 \over (i \omega_n)^2 V} \langle \left[ \rho ({\bf q}, 0), \frac{\rho^{\dag} ({\bf q}, \tau)}{d \tau} \right]_{\tau=0} \rangle
 \\ \nonumber
&-& {1 \over (i \omega_n)^2 V} \int_0^{\beta} d \tau e^{i \omega_n \tau}
\langle T_{\tau} \left[ \frac{d \rho^{\dag} ({\bf q}, \tau')}{d \tau'} \right]_{\tau'=0} \frac{\rho ({\bf q}, -\tau)}{d \tau} \rangle .
\end{eqnarray}
The time derivative of density operator is given by
\begin{eqnarray} \label{eq:eomr}
\frac{d \rho ({\bf q}, \tau)}{d \tau} = [H, \rho ({\bf q}, \tau)]
= {1 \over e} {\bf q} \cdot {\bf j} ({\bf q}, \tau) .
\end{eqnarray}
Substituting into Eq.~\eqref{eq:chi3}, we have
\begin{eqnarray} \label{eq:chi4}
\chi ({\bf q}, i \omega_n) &=& - \left( {q \over i \omega_n} \right)^2 {n_f \over m V}
    - \left( {q \over i \omega_n e} \right)^2 {1 \over V}
\nonumber \\
& & \times \int_0^{\beta} d \tau e^{i \omega_n \tau}
\langle T_{\tau} j_{\parallel}^{\dag} ({\bf q}, \tau) j_{\parallel} ({\bf q}, 0) \rangle ,
\end{eqnarray}
where $n_f = \sum_{\bf k} \langle a_{\bf k}^{\dag} a_{\bf k} \rangle$, and $j_{\parallel} ({\bf q})$ denotes the current along the $\bf q$ direction.
The second term relates to the current-current correlation function in Eq.~\eqref{momdepcond}.
Therefore, after the analytic continuation we get
\begin{eqnarray}
\chi ({\bf q}, \omega) = \left( {q \over \omega} \right)^2
    \left[ {\Pi_{\parallel} ({\bf q}, \omega) \over e^2} - {n_f \over m V} \right] ,
\end{eqnarray}
which finally leads to
\begin{eqnarray}
R_{\chi} ({\bf q}, \omega) = \left( {q \over \omega e} \right)^2 R_{\Pi}^{\parallel} ({\bf q}, \omega) .
\end{eqnarray}

\section{Analytic continuation of {\color{black} the} two-particle Green's function}
\label{sec:app_ac}

Both optical absorption spectrum and Bragg spectrum are two-particle spectral functions.
They can be extracted from the corresponding two-particle Green's functions by solving an integral equation
\begin{eqnarray} \label{eq:ac1}
G (\tau) = - \int_{- \infty}^{\infty} d \omega R (\omega)
\frac{e^{- \tau \omega}}{1 - e^{- \beta \omega}} ,
\end{eqnarray}
where $G(\tau)$ and $R(\omega)$ are assumed to be the two-particle Green's function and its spectrum, respectively (we drop the momentum index for simplicity).
As Eq.~(\ref{eq:ac1}) connects the imaginary time with real frequency, the spectral reconstruction associated with Eq.~(\ref{eq:ac1}) is also known as analytic continuation.
In our calculation, $G(\tau)$ is obtained {\color{black} by a} QMC simulation, and we derive $R(\omega)$ from Eq.~(\ref{eq:ac1}) by using a renormalizing iterative fitting method.
The iteration scheme is originally developed for the one-particle spectral function of {\color{black} an} electron \cite{Ji2004}.
It relies on the sum rule of electronic spectrum, which conserves the total spectral weight through the iteration process.
However, for the two-particle spectrum such a sum rule {\color{black} does not exist}, and the spectral sum is not a conserved quantity.
In this case the spectrum features the following properties:
\begin{eqnarray}
\omega R (\omega) &\ge& 0 , \\
R (-\omega) &=& -R (\omega) .
\end{eqnarray}
Obviously, the two-particle spectrum is anti-symmetric and the total sum of spectral weight equals to zero,
\begin{eqnarray}
\int_{- \infty}^{\infty} d \omega R (\omega) = 0 .
\end{eqnarray}
Therefore, the standard iterative fitting method does not work here.

Nonetheless, because the spectrum is anti-symmetric, we can confine the calculation in the region $0 < \omega < \infty$.
We also can introduce a modified spectral function
\begin{eqnarray} \label{eq:ac2}
\tilde{R} (\omega) = - \frac{R (\omega)}{G (\beta)}
\coth \left( {\beta \omega \over 2} \right) .
\end{eqnarray}
Since $R (\omega)$ is anti-symmetric, on substituting $\tilde{R} (\omega)$ in Eq. (\ref{eq:ac1}), we can rewrite it into
\begin{eqnarray} \label{eq:ac3}
G (\tau) &=& - \int_0^{\infty} d \omega \tilde{R} (\omega)
    \cosh^{-1} \left( {\beta \omega \over 2} \right)
\nonumber \\
& & \times \cosh \left( {\beta - \tau \over 2} \omega \right) G (\beta) .
\end{eqnarray}
Here $\tilde{R} (\omega)$ is just a renormalized form of the original spectral function, but it is easy to see this new function is positive for $\omega>0$, and it satisfies a sum rule
\begin{eqnarray}
\int_0^{\infty} d \omega \tilde{R} (\omega) = 1 ,
\end{eqnarray}
which allows us to solve the integral equation \eqref{eq:ac3} with the iteration algorithm in the regime $0 < \omega < \infty$.
Once $\tilde{R} (\omega)$ is obtained, the original spectral function $R (\omega)$ can be determined from Eq. (\ref{eq:ac2}).

\bibliographystyle{elsarticle-num} 
\bibliography{BEC_optical_spectra}

\end{document}